\documentstyle[12pt]{article}
\begin{document}{
\thispagestyle{empty}

\vskip 2.0cm
{\renewcommand{\thefootnote}{\fnsymbol{footnote}}
\centerline{\large \bf The Folding Funnel Landscape for the Peptide
                       Met-Enkephalin}

\vskip 2.0cm
 
\centerline{Ulrich H.E.~Hansmann$^{1,}$, 
\footnote{\ \ e-mail: hansmann@mtu.edu; to whom all correspondence should be
                                         addressed}
Yuko Okamoto$^{2,}$
\footnote{\ \ e-mail: okamotoy@ims.ac.jp}
and Jose N. Onuchic$^{3,}$
\footnote{\ \ e-mail: jonuchic@ucsd.edu}} 
\vskip 1.5cm
 \centerline{$^1$\it Department of Physics}
\centerline{\it Michigan Technological University} 
\centerline{\it   Houghton, MI 49931-1295, USA}
\vskip 1.5cm
\centerline {$^2${\it Department of Theoretical Studies}} 
\centerline{{\it Institute for Molecular Science}} 
\centerline {{\it Okazaki, Aichi 444-8585, Japan}}
\vskip 1.5cm
\centerline{$^3${\it Department of Physics}}
\centerline{{\it University of California at San Diego}}
\centerline{{\it La Jolla, CA 92093-0319,USA}}

\medbreak
\vskip 3.5cm
 
\centerline{\bf ABSTRACT} \vskip 0.3cm We study the free energy
landscape of the small peptide Met-enkephalin.  Our data were obtained
from a {\it generalized-ensemble} Monte Carlo simulation taking the
interactions among all atoms into account. We show that the free
energy landscape resembles that of a funnel, indicating that
this peptide is a good folder. Our work demonstrates that the energy
landscape picture and folding concept, developed in the context of
simplified protein models, can also be used to describe the folding in
more realistic models.  \vfill} \newpage
 \baselineskip=0.8cm
\noindent
{\bf INTRODUCTION}\\

It is well known that a large class of proteins folds spontaneously
into unique, globular shape \cite{Anf}. However, the mechanism of
protein folding has remained elusive.  To describe many biochemical
processes it may be sufficient to assert that folding occurs on a
time scale no slower than protein biosynthesis, and that the
information required to find the precise three-dimensional shape is
contained in the one-dimensional sequence of the molecule. This simple
description may not be sufficient if the prediction of protein
structure from sequence and the design of truly novel protein-like
molecules are to be achieved. In order to answer the practical
questions of structure prediction and design, it seems that one must
go a considerable distance beyond this phenomenology --- a new
viewpoint may be required.  Such a new viewpoint is emerging from the
analytical and numerical studies of minimal protein models by several
groups (see for example refs.~\cite{DC} -\cite{Onuchic97}). 
Its framework is provided by energy landscape theory and the funnel
concept, which assert that a full understanding of the folding process
requires a global overview of the landscape. The folding landscape of
a protein resembles a partially rough funnel riddled with traps where
the protein can transiently reside. There is no unique pathway but a
multiplicity of convergent folding routes towards the native state
\cite{Bryngelson87}-\cite{Onuchic97}.

The importance of a funnel landscape can be seen by contrasting random
heteropolymeric molecules and proteins.  Both random heteropolymers 
and proteins have an underlying driving force to collapse, and for both
molecules the various competing interactions within the molecule and 
between the molecule and the surrounding solvent lead to a rugged 
energy landscape. However, unlike random heteropolymers, proteins adopt 
well-defined three-dimensional structures because  there is a 
sufficient overall slope of the energy landscape so that the numerous
valleys flow in a funnel toward the native structure.
\cite{DC,LMO}, \cite{BOSW}-\cite{SOW}.

The essence of the funnel landscape idea is competition between the
tendency towards the folded state and trapping due to ruggedness of
the landscape.  This competition is measured by the ratio between the
folding temperature ($T_f$) and the glass temperature ($T_g$). Good
folding protein sequences fold rapidly on minimally frustrated
landscapes with large values of
$T_f/T_g$\cite{Bryngelson87}-\cite{Onuchic97}. Minimally frustrated
sequences not only fold fast at relevant temperatures but are also robust
folders, and therefore only weakly dependent on minor variations of
the folding environment or to mutations. Energy landscape theory
suggests a diversity of folding scenarios that have been explored by
computer simulations of minimalist models (see for example references
in \cite{DC,Onuchic97}, \cite{Thirumalai95}-\cite{Hardin98}) 
and connections to studies of real proteins were
proposed\cite{OWLS,Wolynes96}.

A  quantitative understanding of how the general parameters of
the landscape relate to particular properties of a protein's sequence
or final folded topology requires detailed molecular (and therefore
less coarse-grained) calculations of folding free energy landscapes.
 Unfortunately, simulations of more realistic models of proteins where
the interactions among all atoms in a protein are taken into account
have been notoriously difficult (see for example Ref.~\cite{Vas} for a
review).  These studies involve extremely intensive numerical
calculations, and thus the number and size of systems that can be
explored is limited. In addition, it is difficult for such simulations
to provide a direct view of protein folding dynamics.  Because of the
rough energy landscape, simulations based on canonical Monte Carlo or
molecular dynamics techniques will get trapped at low temperatures in
one of the multitude of local minima separated by high energy
barriers. Hence, there is always the danger that only small parts of
configuration space are sampled and physical quantities cannot be
calculated accurately.

New numerical approaches have been developed to deal with these sampling
difficulties. {\it Generalized-ensemble} techniques (for a recent
review, see, for instance, Ref.~\cite{OurReview98}) like
multicanonical algorithms \cite{MU} and simulated tempering
\cite{ST} allow an efficient sampling of low-energy configurations, and
calculation of accurate low-temperature thermodynamic quantities
became feasible.  The first application of one of these techniques to
the protein-folding problem is given in Ref.~\cite{HO}. 
By comparing with recent experiments the usefulness of the approach was 
extensively tested and demonstrated in  a study of helix-coil transitions of
homo-oligomers of nonpolar amino acids \cite{HO95a,HO98c} and, more recently,
of the C-peptide of ribonuclease A \cite{HO97d,HO98d}.
A numerical comparison of three different generalized-ensemble algorithms 
can be found in  Ref.~\cite{HO96b}.

Even with these new and sophisticated techniques, there are  still many
concerns and difficulties in fully exploring the energy landscape of
even medium-sized proteins, and for this reason we have  restricted
the analyses to only small peptides in this paper.

In general, such peptides are too small to fold and therefore not
useful as a model for the folding process in larger proteins.
However, there are a few small peptides known which exist at low
temperatures in stable defined conformations.  For one of these
peptides the characteristic temperatures of folding were recently
determined  numerically in Ref.~\cite{HMO97b}. These temperatures
were shifted to lower values than expected for larger proteins, but
otherwise the results of that study  indirectly support the
energy landscape picture and funnel concept. Here, we re-examine our
data to  directly investigate the free energy landscape of the
peptide. In that respect, our work is similar to the earlier work
in ref.~\cite{Brooks}  but uses this new sampling technique.

 Our system of choice is the linear peptide Met-enkephalin which
has the amino-acid sequence Tyr-Gly-Gly-Phe-Met.  The experimental
studies of  this  penta-peptide \cite{xray}-\cite{solv} were mainly
motivated by its biological significance as a
neurotransmitter. Compact and defined structures were observed for
membrane bound molecules \cite{nmr}, in crystals \cite{xray} or
organic solvents. However, studies of  this  peptide in aqueous
solution indicate that it exists at room temperature  as an
ensemble of extended coil structures with low frequency of folded
structures \cite{solv}.

 Since the energy landscape is a function of a large number of
degrees of freedom, it is obviously impossible to keep track of all
coordinates. Hence, to actually  observe the folding funnel of
the peptide,  one has to study a projection of the landscape
onto a set of suitable and appropriate order parameters.  For our
choice of the order parameters,  we were guided by previous
numerical simulations of Met-enkephalin
\cite{HMO97b},\cite{LS}-\cite{MHO}. Using the ECEPP/2 force field \cite{EC3}, 
it was
shown in a recent article \cite{HMO97b} that Met-enkephalin undergoes
a transition between extended and compact structures
at a temperature $T_{\theta} = 295 \pm 20$ K.
Above that temperature,  the frequency of compact structures rapidly
decreases while it increases below $T_{\theta}$.  Hence, our first
order parameter is the volume allowing us to distinguish between
compact and extended conformations.  In Ref.~\cite{HMO97b} it was also
shown that by further lowering the temperature the peptide encounters
a second transition.  Below $T_f = 230 \pm 30$ K, the
occupation of the ground-state conformations increases rapidly while
it decreases for values of $T$ above $T_f$.  The ensemble of low-temperature
conformations was studied by various groups
\cite{HO},\cite{LS}-\cite{HO94_3},\cite{EH96} and the numerical results 
compared
with the experimental findings \cite{Wils}.  These studies agree in
that there are two major groups of well-defined compact structures
which are characterized (and stabilized) by specific hydrogen bonding
 patterns.  In Fig.~1 we show a sketch of the two structures.
Structure A is the 
ground-state conformation in ECEEP/2 and  has a Type II' $\beta$-turn
between the second and last residue, stabilized by two possible
hydrogen bonds.  The structure B, the second-lowest energy state, is  
characterized by  hydrogen bond between Tyr-1 and Phe-4 
resulting in a Type-II $\beta$-turn between the first and
fourth residue.  
  We remark that at  higher energies
 (in ECEPP more than 2 kcal/mol above the ground state) conformations with
 a $\gamma$-turn and  those with hydrogen bonding between the first and 
 last residue were also observed \cite{OK,MHO}. 
Hence, we choose as further order parameters the overlap
with the ground state (structure A) and the second-lowest-energy
state (structure B), respectively, which allows us to distinguish between
the various compact low-energy conformations.

In the following sections we first review  our 
simulation method and the ECEPP force field and show afterwards 
how the above order parameters can be defined and measured. 
Our results are then presented and we finish with our conclusions.\\
 
\noindent
{\bf SIMULATION TECHNIQUES}\\
The generalized-ensemble technique  utilized in this article was first
introduced in Refs.~\cite{H97a,HO96d}. In this algorithm, configurations are
updated according to the following probability weight: 
\begin{equation}
w(E) = \left(1+ \frac{\beta (E-E_0)}{n_F}\right)^{-n_F}~,
\label{eqwe}
\end{equation}
where $E_0$ is an estimator for the ground-state energy, $n_F$ is
the number of degrees of freedom of the system, and $\beta = 1/k_BT$
is  the inverse temperature   ($k_B$ is the
Boltzmann constant and $T$ the temperature of the system). 
Note that this weight is a
special case of the weights used in Tsallis generalized mechanics
formalism \cite{Tsa} (the Tsallis parameter $q$ is chosen as  $q = 1 + 1/n_F$).
In the low-energy region 
(where $\frac{\beta (E-E_{0})}{n_F} \ll 1$),
the new weight reduces to the canonical
Boltzmann weight $\exp (- \beta E)$. 
 On the other hand, high-energy regions are no
longer exponentially suppressed but only according to a power law,
which enhances excursions to high-energy regions. We remark that the
calculation of the weight is much easier than in other generalized-ensemble
techniques, since it requires one to find only an estimator for the 
ground-state energy $E_0$, which can be done by a procedure described
in Ref.~\cite{HO96d}.

As in the case of other generalized ensembles,  we can
use the reweighting techniques \cite{FS} to construct canonical distributions
at various temperatures.  This is because
the simulation by the present algorithm samples a large range of 
energies. The thermodynamic average of any physical quantity $\cal{A}$
can be calculated over a wide temperature range by
\begin{equation}
<{\cal{A}}>_T ~=~ \frac{\displaystyle{\int dx~{\cal{A}}(x)~w^{-1}(E(x))~
                 e^{-\beta E(x)}}}
              {\displaystyle{\int dx~w^{-1}(E(x))~e^{-\beta E(x)}}}~,
\label{eqrw}
\end{equation}
where $x$ stands for configurations.\\
\\

\noindent
{\bf FORCE FIELDS}\\
For our simulations we used the ECEEP/2 force field \cite{EC3} in which 
 the potential energy function $E_{tot}$ is given by the sum of
the electrostatic term $E_{C}$, 12-6 Lennard-Jones term $E_{LJ}$, and
hydrogen-bond term $E_{HB}$ for all pairs of atoms in the peptide
together with the torsion term $E_{tor}$ for all torsion angles:
\begin{eqnarray}
E_{tot} & = & E_{C} + E_{LJ} + E_{HB} + E_{tor},\\
E_{C}  & = & \sum_{(i,j)} \frac{332q_i q_j}{\epsilon r_{ij}},\\
E_{LJ} & = & \sum_{(i,j)} \left( \frac{A_{ij}}{r^{12}_{ij}}
                                - \frac{B_{ij}}{r^6_{ij}} \right),\\
E_{HB}  & = & \sum_{(i,j)} \left( \frac{C_{ij}}{r^{12}_{ij}}
                                - \frac{D_{ij}}{r^{10}_{ij}} \right),\\
E_{tor}& = & \sum_l U_l \left( 1 \pm \cos (n_l \chi_l ) \right).
\end{eqnarray}
Here, $r_{ij}$ (in \AA) is the distance between the atoms $i$ and $j$, and
$\chi_l$ is
the torsion angle for the chemical bond $l$. Bond lengths and bond angles
are  fixed at experimental values, 
leaving the dihedral angles $\phi,~\psi,~\omega$, and $\chi$
as independent variables. We further fix the peptide bond angles $\omega$
 to their common value $180^{\circ}$, which
 leaves us with 19 torsion angles ($\phi,~\psi$, and $\chi$) as independent
degrees of freedom (i.e., $n_F = 19$).
In our simulations we did not  explicitly include the interaction
of the peptide with the solvent and set the dielectric constant $\epsilon$ 
equal to 2. However, we do expect some implicit solvent effect, since 
 the various parameters ($q_i,A_{ij},B_{ij},C_{ij},
D_{ij},U_l$, and $n_l$) for the energy function were
determined by  minimization
of the potential energies of the crystal lattices of single amino acids,
i.e., not in a vacuum.
We remark that the computer code KONF90 \cite{KONF} which we used in our
simulation relies on a different convention for the implementation of the
 ECEPP parameters (for example, $\phi_1$ of ECEPP/2 is equal to
 $\phi_1 - 180^{\circ}$ of KONF90).  Therefore, our energy values are
slightly different
from those of the original implementation
of ECEPP/2.\\
\\
{\bf TECHNICAL DETAILS}\\
It is known from our previous work that the ground-state conformation
for Met-enkephalin has the
KONF90 energy value
$E_{GS} = -12.2$ kcal/mol \cite{HO94_3}. We therefore set $E_0 = -12.2$
kcal/mol, $T = 50$ K (or, $\beta = 10.1$ $[\frac{1}{{\rm kcal}/{\rm mol}}]$) 
and $n_F =19$ in our probability weight factor in Eq.~(\ref{eqwe}).
Our simulation was started from a completely random initial
conformation (Hot Start) and  one Monte Carlo sweep updates every torsion angle
of the peptide once.
All thermodynamic quantities were then  calculated from
 a single production run of 1,000,000 MC sweeps which followed 10,000
sweeps for thermalization. At the end of every fourth sweep we stored
the energies of the conformation and our three ``order parameters''
(the corresponding volume, the overlap $O_A$ of the conformation with
the (known) ground state (structure A) and the overlap $O_B$ of the
 conformation with conformer B.  Here, we approximate the volume
of the peptide by its solvent excluded volume (in \AA$^3$) which is
calculated by a
 variant \cite{MO} of the double cubic lattice method \cite{ELASS}.  
Our definition of the overlap, which measures how much a given conformation
resembles a reference state, is given by
\begin{equation}
O(t) = 1 -\frac{1}{90~n_F} \sum_{i=1}^{n_F} |\alpha_i^{(t)}- \alpha_i^{(RS)}|~,
\label{eqol}
\end{equation}
where $\alpha_i^{(t)}$ and $\alpha_i^{(RS)}$ (in degrees) stand for 
the $n_F$ dihedral angles of the conformation at $t$-th Monte Carlo sweep 
and the  reference state conformation, respectively. Symmetries
for the side-chain angles were taken into account and the difference
$\alpha_i^{(t)}- \alpha_i^{(RS)}$ was always projected into the interval
$[-180^{\circ},180^{\circ}]$. Our definition  guarantees that we have 
\begin{equation}
0 \le ~<O>_T~ \le 1~.
\end{equation}
We remark that the average overlap $<O>_T$ approaches its limiting
value zero (for $T \rightarrow \infty$) only very slowly as the
temperature increases. For instance, at $T=1000$ K we found for the
overlap with the ground state still  has an average value of
$<O_A> \approx 0.3$. This is because $<O>_T ~= 0$ corresponds to a
completely random distribution of dihedral angles which is
energetically highly unfavorable  due to the steric hindrance of
both main and side chains.  Note the obvious limit: $O_A \rightarrow
1$, as $T \rightarrow 0$.

Since large parts of the configuration space are sampled by our method,
the use of the reweighting techniques \cite{FS} is justified to 
calculate thermodynamic quantities over a wide range of temperatures
by Eq.~(\ref{eqrw}). Examples are the average potential energy $<E>(V)$,
the entropy 
\begin{equation}
S(V) = <E>(V)/k_B T + \log P(V)
\label{eqSV}
\end{equation}
 (where $P(V)$ is the 
probability to find a conformation with the volume $V$), and 
similar quantities defined as functions of the  
two overlaps $O_A$ and $O_B$. We normalized all the above 
quantities in such a way that they are zero for the ground state.

The above defined quantities allow only to study a projection of the
energy landscape into one dimension. To get a more detailed 
picture, we also explored for various temperatures the free energies
\begin{equation} 
G(O_A,O_B) = -k_B T \log P(O_A,O_B) 
\end{equation}
and 
\begin{equation} 
G(O_A,V) = -k_B T \log P(O_A,V)~, 
\end{equation}
where again $P(O_A,O_B)$ and $P(O_A,V)$ are respectively the
 probabilities to find a peptide conformation with values $O_A$, $O_B$
and $O_A$, $V$.  We chose the normalization so that the lowest value
of $G(O_A,O_B)$ or $G(O_A,V)$ is set to zero for each temperature.

Finally, in order to monitor the number of states in which the
peptide can be found at temperature $T$, we also calculated the 
(unnormalized) entropy
\begin{equation}
S(T) = <E>(T)/k_BT -  \log Z(T)~,
\label{ST}
\end{equation}
where $Z(T)$ is the estimate of the partition function of the system 
as calculated
from our data by the reweighting techniques. This quantity allows one to 
estimate the  glass temperature $T_g$, since the number of possible states
should decrease drastically after entering the glassy phase.\\

\noindent
{\bf THE MET-ENKEPHALIN FUNNEL LANDSCAPE}\\
The energy landscape for a folding protein strongly depends on
temperature. For some temperatures non-specific trapping may not be a
problem but the situation may completely reverse for other
temperatures. Therefore, when exploring the folding landscape for this
peptide, the simulations should take place at relevant temperatures, for 
otherwise one is not able to obtain a reliable picture of the folding
mechanism. Hence, we have concentrated our analyses on four
temperatures. The first one, $T=1000$ K was chosen to probe the 
high-temperature  regime where the peptide is fully unfolded and
mostly in an extended form. In some early work~\cite{HMO97b}, some of
us have identified $T=300$ K as the collapse temperature $T_{\theta}$
and $T=230$ K as the folding temperature $T_f$. The last temperature,
$T=150$ K, was chosen to study the low temperature behavior of the
peptide where the glassy behhavior is observed.  

Fig.~2a displays the average potential energy $ <E>(V)$ 
as a function of the volume of the peptide for the four chosen temperatures.
 The plot shows that configurations with small volume
($< 1400$ \AA$^3$) have essentially the same  energy.  Above that value
of the volume the energy increases with the volume.  The increase is
only gradual for high temperatures, but very steep below $T_{\theta}$.
Similar results were found for the entropy $ S(V)$, displayed for the
same four temperatures in Fig.~2b. The number of states varies
slightly below a certain threshold but increases rapidly above that
value. The steepness of that increase is again a function of
temperature below $T_{\theta}$. The two plots indicate that one can
distinguish between two regimes: compact structures which have similar
energies and entropy and extended structures which are entropically
favored but energetically disfavored.

Fig.~3a (3b) shows the average energy (entropy) 
as a function of the overlap with structure A (the ground state
in ECEPP/2).  For the interpretation of the plots one has
to keep in mind that the overlaps approach zero very slowly with
increasing energy or temperature. Results from our simulation are 
only trustworthy for temperatures $T\le 1000 $ K where the overlap
function has average value of $\approx 0.3$. Hence, in Figs.~3a and 3b 
 the results are only reliable for values of the overlap function
above 0.3.

At high temperatures, the dependence of $<E>(O_A)$ on the overlap
$O_A$ is essentially a monotonous function where the energy decreases
with increasing value of the overlap. Below the collapse temperature
$T_{\theta}$, however, the curves show a minimum at $O_A \approx
0.5$. This is the value $O_A$ of Conformation B (the exact value being
0.46), and this minimum is therefore an indication of a basin of
attraction to  Conformer B.
However,  the overall slope  of $<E>(O_A)$ shows that Conformer A
is favored.  On the other hand, 
the entropy as a function of the overlap strongly varies with
decreasing overlap for all but the lowest temperature. The absolute
values of the entropy decreases with temperature. At $T=150$ K,
however, $S(O_A)$ stays essentially constant over the whole range of
values of the overlap once this value is only a little different from
1. This  means that at this temperature the number of states
differs little with the overlap and therefore indicates the onset of
glassy behavior.  Similar pictures hold when one prints energy and
entropy differences as function of the overlap with Conformation B.
This kind of behavior has been predicted from simulations of
minimalist models and now they have been confirmed for simulations of
a real peptide.

As the number of available states  gets reduced with the decrease
of temperature, the possibility of local trapping increases
substantially. In the thermodynamic limit, the system would  be
trapped in these local traps for an infinite time as expected 
for a glass transition. To determine  the glassy behavior for a
finite system is a more subtle problem. In the folding problem, glassy
behavior is associated when the residence time in some local traps
becomes of the order of the folding event. Folding dynamics is now
non-exponential since different traps have different escape 
times~\cite{REF1,Nymeyer98}. For temperatures above the glass
transition temperature $T_g$, the folding dynamics is exponential and
a configurational diffusion coefficient average the effects of the
short lived traps \cite{SOW,Baker97,Plotkin98}. It is expected
that for a good folder the temperature, $T_g$, where glass behavior
sets in, has to be significantly lower than the folding temperature
$T_f$, i.e. a good folder can be characterized by the 
relation \cite{Bryngelson87}
\begin{equation}
\frac{T_f}{T_g} > 1~.
\end{equation}
 Our plots of $\Delta S(O_A)$ indicate that we indeed have $T_g
< T_f$.  This is also supported by Fig.~4 where the (unnormalized)
entropy of the molecule is represented as a function of
temperature. This plot clearly shows the rapid decrease of entropy
once $T$ reaches values smaller than $T_g$. As one would expect from
such a small molecule, it is  impossible to determine a precise
value for $T_g$ from the plot, but clearly the onset of glassy
behavior happens at temperatures well below the folding temperature
$T_f=230$ K. 
In Ref.~\cite{HMO97b} it was already pointed out that these
results are also consistent with an alternative characterization of
folding properties. Thirumalai and collaborators \cite{CTh,KTh} 
have conjectured
that  the kinetic accessibility of the native conformation can be classified
by the parameter  
\begin{equation}
\sigma = \frac{T_{\theta} - T_f}{T_{\theta}}~,
\label{sig}
\end{equation}
i.e., the smaller $\sigma$ is, the more easily a protein can fold.
With our values for $T_{\theta}$ and $T_f$, we have for Met-enkephalin
$\sigma \approx 0.2$.  Here, we have taken the central values:
$T_{\theta} = 295$ K and $T_f = 230$ K.
This value of $\sigma$ implies that our peptide has reasonably
good folding properties
according to Refs.~\cite{CTh} and \cite{KTh}. Hence, we see that there is a
strong correlation between the folding criterion ($T_f/T_g > 1$) proposed by
Bryngelson and Wolynes \cite{Bryngelson87} and this one.

The main goal of this article is to depict the folding funnel of 
Met-enkephalin. Since it is not feasible to plot the free energy $G$ 
as a function of all three order parameters, one has to plot
$G$ as  a function of a suitable combination of the three relevant 
order parameters of the molecule. We chose 
to plot the free energy $G(V,O_A)$ as a function of volume $V$ and
overlap $O_A$ with the known ground state (Conformer A) and 
$G(O_A,O_B)$ as a function of the overlap with the ground state ($O_A$)
and with Conformer B ($O_B$). Again we study these quantities for
temperatures $T=1000$ K, $T=T_{\theta}=300$ K, $T=T_f=230$ K,
and $T=150$ K. 
They are shown in Fig.~5a-d and Fig.~6a-d, respectively.

In Fig.~5a we show the free energy landscape as a function of volume
and overlap with the known ground state (structure A)  at the
high-temperature region ($T=1000$ K). Here, (as in the other free
energy plots)  we normalized the free energy in such a way that
its observed minimum is set to zero. In the contour plots, the
contour lines mark multiples of $k_B T$  (therefore different for
different temperatures but appropriate to understand the folding
mechanism). We see that the free energy has its minimum at large
volumes ($\approx 1470$ \AA$^3$) and values of the overlap $O_A
\approx 0.3$. Small volumes and larger values of the overlap are
suppressed by many orders of $k_B T$. Hence, extended random coil
structures are favored at this temperature.  The picture changes
dramatically once we reach the collapse temperature $T_{\theta}$,
shown in Fig.~5b. At this temperature a large part of the $V$-$O_A$
space can be sampled in a simulation. The contour plot shows that
regions with both small and large volumes and almost all values of
$O_A$ lie within the 2 $k_BT$ contour.  This indicates that at
this temperature the cross over between extended and compact
structures happens with a small thermodynamic barrier between them.
By lowering the temperature to $T_f =230 K$ (determined in
ref.~\cite{HMO97b}), we now observe strong evidence for a funnel-like
landscape (Fig.~5c).  At this temperature the drive towards the native
configuration is dominant and no long-lived traps exist.
There is clearly a
gradient towards the ground-state structure ($O_A \approx 1$), but
other structures with similar volume (characterized by values of $O_A
\approx 0.5$) are only separated by free energy barriers of order 1
$k_BT$. Below this temperature we
expect that the ground state is clearly favored thermodynamically and
separated from other low energy states by free energy barriers of many
orders of $k_B T$. This can be seen in Fig.~5d where at $T=150 K$
where other low energy states have free energies of 3 $k_BT$ higher
than the ground state and are separated by an additional barrier of 2
$k_BT$.

The above picture is supported by the plots for the free energy as a
function of both the overlap $O_A$ with the ground state and the
overlap $O_B$ with structure B. Fig.~6a shows again the high-temperature 
situation. The free energy has its minimum at small values
of the overlap indicating that both conformers appear with only very
small frequency at high temperature. At $T=300 K$, the collapse
temperature, again a large part of the space of possible
configurations (characterized by values of $O_A$ and $O_B$) lies
within the $2 k_BT$ contour as is clear from Fig.~6b. At the folding
temperature $T_f = 230$ K a funnel in the energy landscape appears
with a gradient towards the ground state, but Fig.~6c shows that there
are various other structures, the most notable of which is Conformer B
(where $O_B \approx 1$), with free energies 3 $k_BT$ higher than the
ground-state conformation but separated from each other and the ground
state only by free energy barriers less than 1 $k_BT$.  No other
long-lived traps are populated. Hence, the funnel at $T_f$ is
reasonably smooth.  Folding routes  include direct conversion
from random-coil conformations into Conformer A or some short
trapping in Conformer B region before reaching Conformer A region, 
but at the folding
temperature it is possible to reach the ground state from any
configuration without getting  kinetically trapped. 
Kinetic Monte Carlo runs at a fixed temperature ($T=230$ K) 
were performed and confirmed this picture
(data not shown). We observed that some of the runs went directly from the
unfolded state to the ensemble of folded conformations in state A,
while in other 
runs  trapping at state B occurred first before folding into the
ground-state structure. The kinetic
runs therefore support our observation
that Met-enkephalin is a good folder and that $T_f > T_g$.  
Fig.~6d shows the situation for $T= 150$ K where we expect onset of glassy
behavior.  Again one sees a  funnel-like bias toward the ground
state, however, the funnel is no longer smooth and the free energy
landscape is rugged. Free energy barriers of many $k_BT$ now separate 
different regions and would act as  long-lived kinetic traps in a
canonical simulation rendering folding at this temperature extremely
difficult.\\

\noindent
{\bf CONCLUSIONS}\\
In this article we have studied the free energy landscape of the
peptide Met-enkephalin, using generalized-ensemble
techniques. Although the peptide is rather small, the obtained free
energy landscape  shows a funnel towards the ground state 
even for temperatures well below the folding temperature $T_f$. It
was shown that glassy behavior appears only well below this
temperature and that the peptide is a good folder at $T_f$.  Our
results demonstrate that the energy landscape picture and funnel
concept of folding  that have been demonstrated in the past for minimalist
models are much more general and hold for this peptide when simulated
with an all-atom force field.\\

\vspace{0.5cm}
\noindent
{\bf Acknowledgments}: \\
Our simulations were  performed on computers 
of the Institute for Molecular Science (IMS), Okazaki,
Japan.
This work is supported by a Grant-in-Aid for Scientific Research from the
Japanese Ministry of Education, Science, Sports and Culture. Work in
San Diego was also supported by the National Science Foundation
(Grant \# 96-03839). Work in Houghton was supported by funds from
Michigan Technological University.  JNO was a visitor at the IMS when
part of this work was performed and he thanks Y. Tanimura for his
hospitality during his stay in Japan.



\newpage
{\large \bf Figures:}\\
\begin{itemize}
\item{Fig.~1:} Backbone structures of the two dominant low-energy 
               structures with their
               characteristic hydrogen bonding. The figures were 
               created with RasMol \cite{Rasmol}. Fig.~1a displays conformer
               A, the ground state in ECEPP/2 (with a KONF90-energy of
               $-12.2$ kcal/mol). Conformer B in Fig.~1b is the local
               minimum with the second lowest potential energy ($-11.0$
               kcal/mol in KONF90).
\item{Fig.~2:}  a) Potential energy  $< E> (V)$ (in kcal/mol) and
                b) entropy  $ S (V)$ (as defined in Eq.~\ref{eqSV}) 
                   as a function of the 
                   volume (in \AA$^3$) of the peptide for various temperatures
                   (in K).
                We chose units where for the ground state  $<E>(V) =0$ and
                $S(V)=0$.
\item{Fig.~3:}  a) Potential energy  $ <E> (O_A)$ (in kcal/mol) and
                b) entropy  $ S (O_A)$ as a function of the 
                overlap $O_A$ (defined in the text) for 
                various temperatures (in K).
                We chose units where for the ground state  $<E>(O_A) =0$ and
                $S(O_A)=0$.
\item{Fig.~4:}  Unnormalized entropy $S(T)$ (as defined in Eq.~\ref{ST}) 
                as a function of temperature (in K).
\item{Fig.~5:}  Free energy $G(V,O_A)$ (in kcal/mol) as a function of both 
                peptide volume $V$ (in \AA$^3$) and overlap $O_A$ (as 
                defined in the text)
                for a) $T = 1000$ K, b) $T=300$ K, c) $T=230$ K, and d)
                $T=150$ K. Both the free energy surface and the
                 contour plot are shown. The contour lines are multiples of
                $k_BT$. $G(V,O_A)$ was normalized such that
                 $\min(G(V,O_A)) = 0$.
\item{Fig.~6:}  Free energy $G(O_A,O_B)$ as a function of both 
                overlaps $O_A$ and $O_B$ (as defined in the text)
                for a) $T = 1000$ K, b) $T=300$ K, c) $T=230$ K, and d)
                $T=150$ K. Both the free energy surface and the
                contour plot are shown. The contour lines are multiples of
                $k_BT$. $G(O_A,O_B)$ was normalized such that
                $\min(G(O_A,O_B)) = 0$.
\end{itemize}

\begin{thebibliography}{(00)}
\bibitem{MO} The program for the calculation of solvent excluded volume was
             written by M.~Masuya and will be described in detail  elsewhere.
\end{thebibliography}
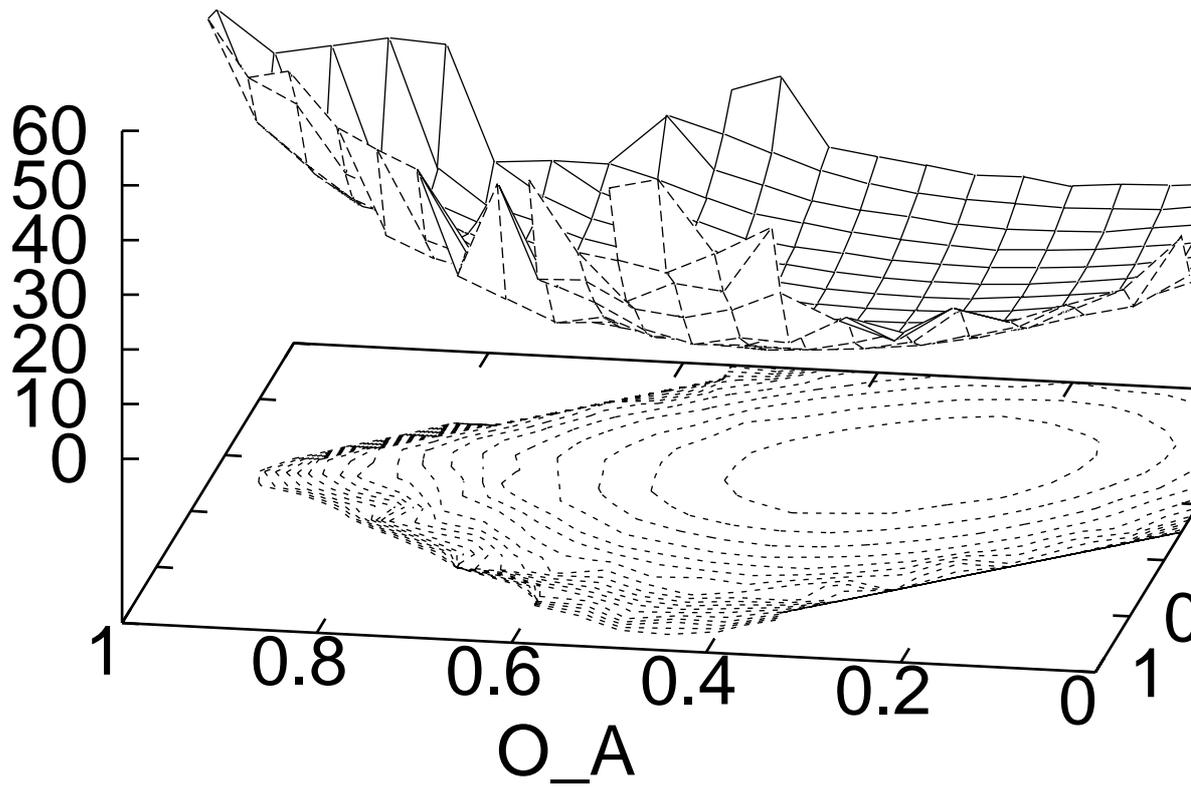
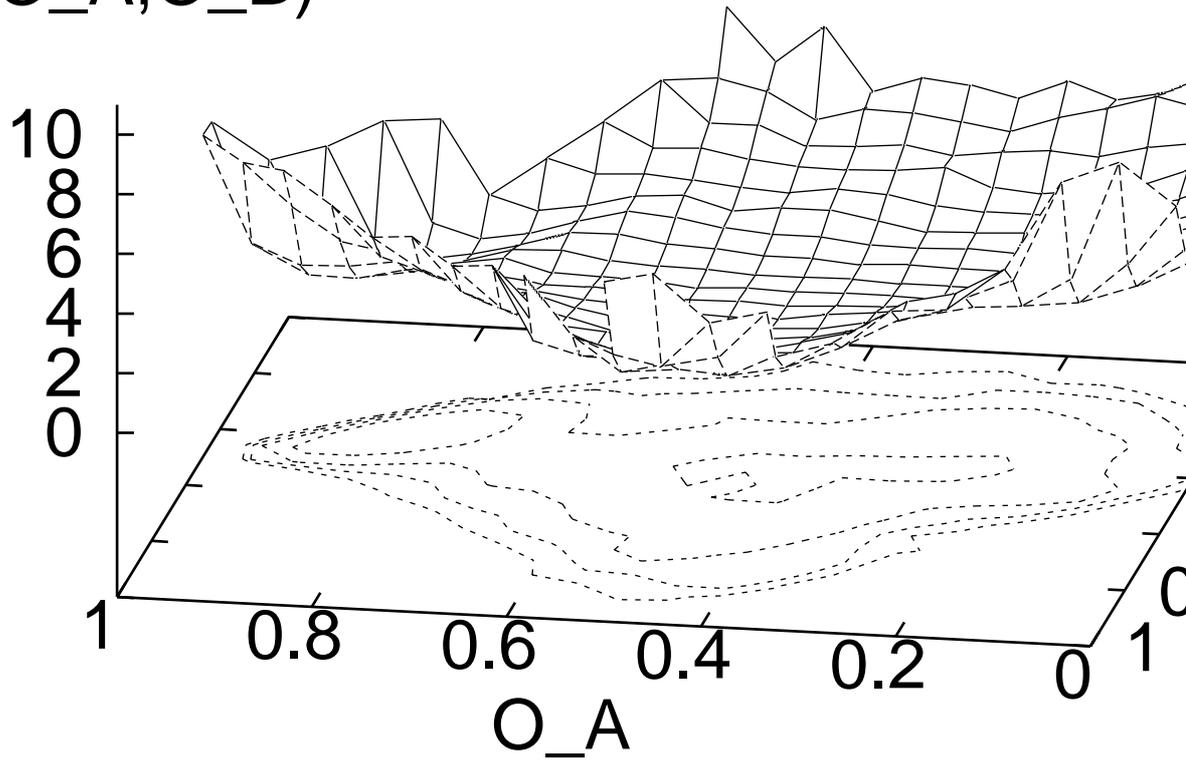
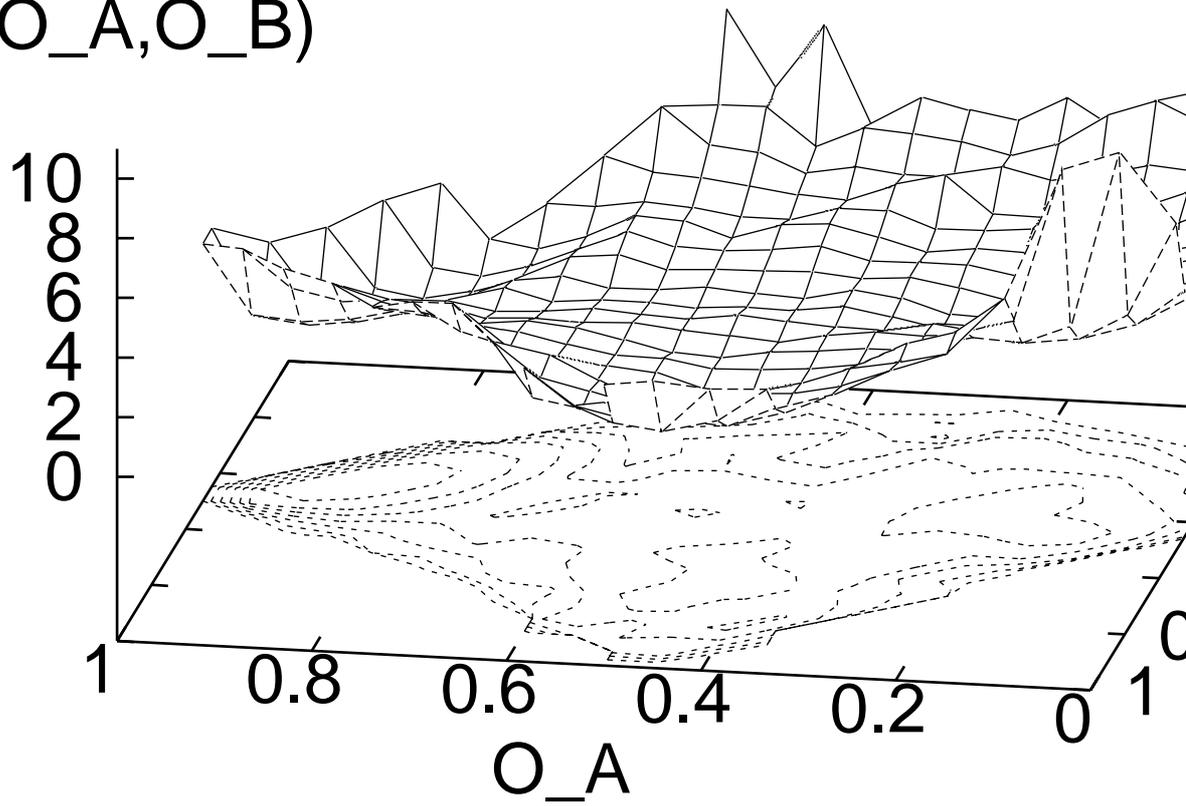
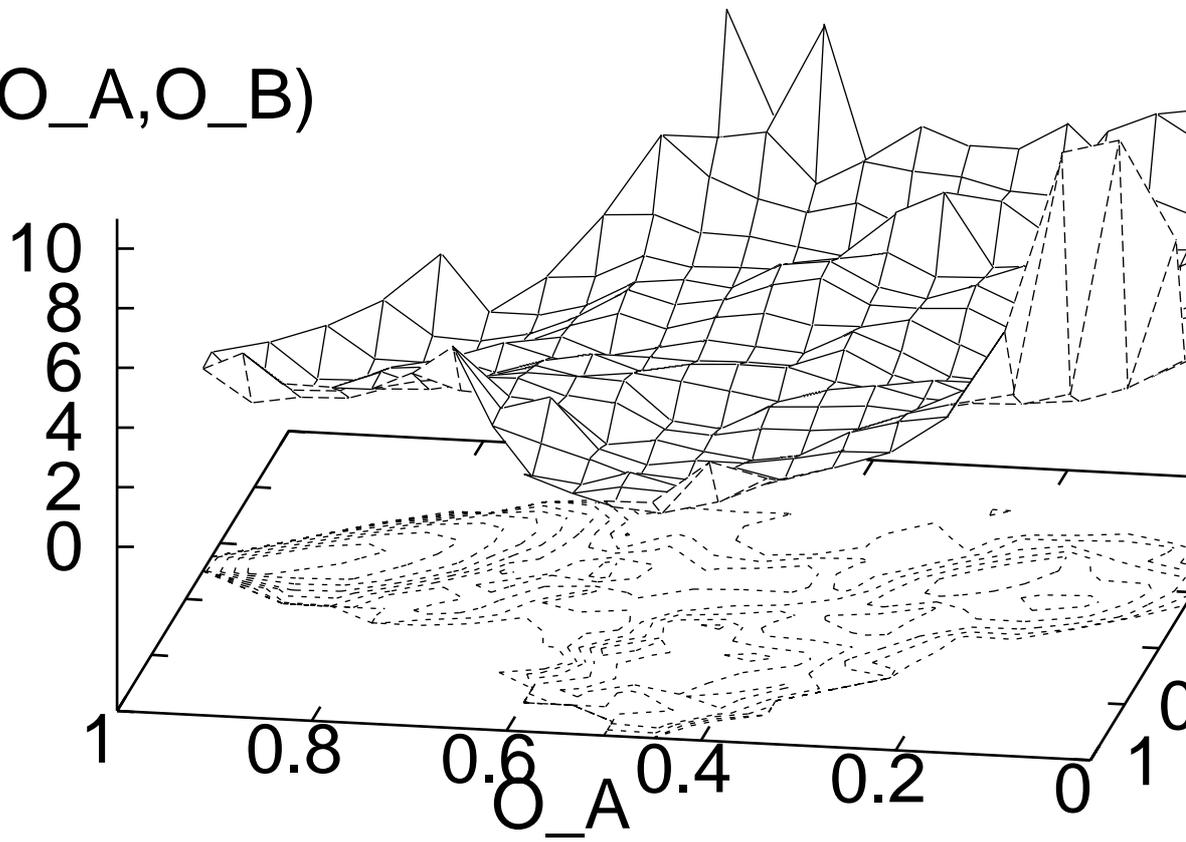
\end{document}